\documentclass[a4paper,12pt]{article}
\usepackage{amsmath}
\usepackage{amsfonts}
\usepackage{amssymb}
\usepackage{latexsym}
\usepackage{epsfig}
\usepackage{graphicx}
\usepackage{oldgerm}
\usepackage{theorem}

\def\R{\mathbb{R}}
\def\N{\mathbb{N}}

\def\r{{\bf r}}
\def\x{{\bf x}}
\def\y{{\bf y}}

\def\1{{\bf 1}}

\def\cN{{\cal N}}

\theorembodyfont{\itshape}

\newcommand{\SSC}[1]{\section{#1}\setcounter{equation}{0}}

\begin{document}

\title{\bf Stochastic model showing a transition
to self-controlled particle-deposition state
induced by optical near-fields}

\author{Kan Takahashi
\footnote{
Department of Physics, Faculty of Science and Engineering, Chuo University, 
1-13-27 Kasuga, Bunkyo-ku, Tokyo 112-8551, Japan;
e-mail:k-takahashi@phys.chuo-u.ac.jp
}, \,
Makoto Katori
\footnote{
Department of Physics, Faculty of Science and Engineering, Chuo University, 
1-13-27 Kasuga, Bunkyo-ku, Tokyo 112-8551, Japan;
e-mail:katori@phys.chuo-u.ac.jp
}, \,
Makoto Naruse
\footnote{
National Institute of Information and Communications Technology, 
4-2-1 Nukui-kita, 
Koganei, Tokyo 184-8795, Japan
}, \,
Motoichi Ohtsu
\footnote{
Nanophotonics Research Center, Graduate School of Engineering, The University of Tokyo, 
2-11-16 Yayoi, Bunkyo-ku, Tokyo 113-8656, Japan
}}
\date{28 July 2015}
\pagestyle{plain}
\maketitle
\begin{abstract}
We study a stochastic model for the self-controlled particle-deposition
process induced by optical near-fields.
This process was experimentally realized by Yukutake et al.
on an electrode of a novel photovoltaic device
as Ag deposition under light illumination,
in which the wavelength of incident light
is longer than the long-wavelength cutoff of the materials
composing the device. 
Naruse et al. introduced a stochastic cellular automaton model
to simulate underlying nonequilibrium processes
which are necessary to formulate unique granular Ag film
in this deposition process.
In the present paper, we generalize their model 
and clarify the essential role of optical near-fields 
generated on the electrode surface.
We introduce a parameter $b$ 
indicating the incident light power per site
and a function representing the resonance effect
of optical near-fields depending on the Ag-cluster size
on the surface.
Numerical simulation shows a transition
from a trivial particle-deposition state to a nontrivial
self-controlled particle-deposition state at a critical value $b_{\rm c}$,
and only in the latter state optical near-fields are effectively generated.
The properties of transition in this mesoscopic surface model in nonequilibrium
are studied by the analogy of equilibrium phase transitions
associated with critical phenomena
and the criteria of transition are reported.
\end{abstract}
                             
\SSC{Introduction}
\label{sec:introduction}

Nanophotonics, which investigates light-matter interactions 
on the nanometer scale, has been intensively studied from 
a variety of aspects ranging from fundamental interests, 
such as atom and optical near-field interactions \cite{Ohtsu08,Tojo05,Ohtsu14}, 
to applications including environment and energy \cite{Franzl04,Yukutake10}, 
healthcare \cite{Pistol08}, solid-state lighting \cite{Kawazoe11}, information 
and communications technologies \cite{Naruse13,Naruse13b}, among others. 
Precision control of the geometrical features, such as the size, 
layout, morphology, at the nanometer scale are important in realizing 
valuable functionalities \cite{Ohtsu08,Ohtsu14}. 
In this context, the nanophotonics fabrication-principles and 
techniques have been providing interesting and important
light-assisted, self-controlled nanostructures; 
nanoparticle array formation \cite{Yatsui05}, light emission 
from indirect-transition-type semiconductors (such as silicon) \cite{Kawazoe11}, 
appearances of photosensitivity at wavelengths 
longer than the long-wavelength cutoff \cite{Yukutake10}, etc.  
In these light-assisted material formations, while elemental 
physical processes indeed involve optical near-field interactions 
at the nanometer-scale, the system is open to environment 
and thus is accompanied with energy flow and environmental fluctuations. 
For these systems, stochastic modeling \cite{MD99,MKL09} is very useful in order to obtain 
deeper understandings of the underlying physical processes as well 
as to gain design principles for future devices and systems \cite{Ohtsu14}. 
In the present paper, we develop the cellular automaton model \cite{CD98}
proposed by Naruse et al. \cite{Naruse11}, 
which was introduced to simulate self-controlled pattern formation of Ag film
reported by Yukutake et al. \cite{Yukutake10} on the electrode of 
their photovoltaic device.

\begin{figure}
\includegraphics[width=0.6\linewidth]{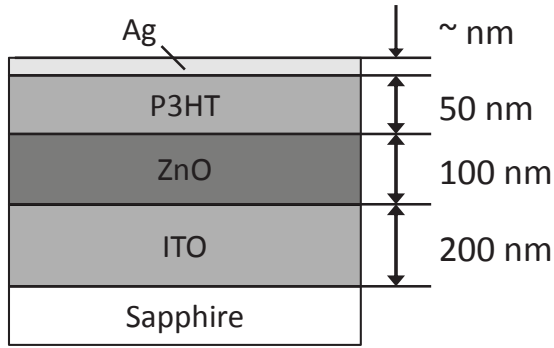}
\caption{Cross-sectional structure of the photovoltaic device
of Yukutake et al.\cite{Yukutake10}.
}
\label{fig:device1}
\end{figure}

\begin{figure}
\includegraphics[width=0.6\linewidth]{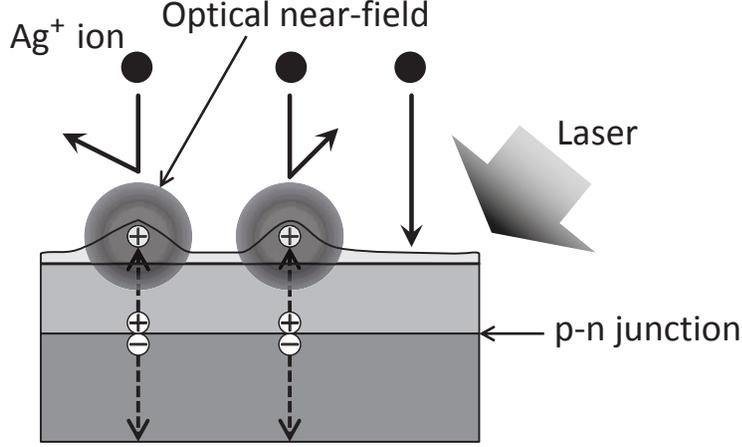}
\caption{Under light illumination, an optical near-field is locally generated on the 
Ag surface, which induces a coherent phonon at the p-n junction.
Then, this generates an electron-hole pair at the p-n junction,
and by the reverse bias voltage, 
the positive hole is attracted to the Ag film on the surface, 
which makes the Ag film positively charged more.
Subsequent deposition of Ag$^{+}$ ion on the electrode is
suppressed by positive charge accumulated in the Ag clusters
on the surface.
}
\label{fig:device2}
\end{figure}

Figure \ref{fig:device1} shows the cross-sectional structure of the photovoltaic device
of poly-(3-hexyl thiophene) (P3HT) and ZnO sandwiched by
Ag and indium tin oxide (ITO) electrodes.
A P3HT layer (about 50 nm thick) and a ZnO layer (about 100 nm thick)
are used as p-type and n-type semiconductors, respectively,
and an ITO film (about 200 nm thick) and an Ag film (a few nanometers thick)
are used as two electrodes. This multi-layered film with an 
area of 30 mm$^2$ is deposited on a sapphire
substrate in series.
At the last stage of the fabrication process of the multi-layered device,
Ag was deposited on the Ag thin film.
The Ag is deposited by radio frequency (RF) sputtering under light illumination
while applying a reverse bias DC voltage, $V_{\rm b}=-1.5$ V, to the 
P3HT/ZnO p-n junction.
The wavelength of the incident light is 670 nm, which is longer than
the cut-off wavelength $\lambda_{\rm c}=570$ nm of P3HT.
Under light illumination, optical near-fields are locally generated on the 
Ag surface. They induce coherent phonons and form a coupled state with them,
which is called a {\it virtual exciton-phonon-polariton} \cite{KSIO00}
or the {\it dressed-photon-phonon (DPP)} \cite{Ohtsu14}.
If the DPP field extends to the p-n junction, the two-step excitation process
of electrons occurs (see \cite{Yukutake10} and Sec.7.2.1 of \cite{Ohtsu14})
and then electron-hole pairs are created at the p-n junction.
As illustrated by Fig. \ref{fig:device2}, by the reverse bias voltage, 
the electrons and holes are separated from each other.
The positive holes are attracted to the Ag film, which make the Ag film be 
positively charged.
It was argued in \cite{Yukutake10,Naruse11} that
due to randomness in Ag deposition process by RF sputtering,
the density of Ag deposits will spatially fluctuate.
Since the optical near-fields are generated by
short-ranged light-matter interaction, 
the optical near-fields will also become spatially inhomogeneous in the film.
In the local area of the Ag surface, where the optical near-field
is effectively induced, more positive holes are generated and 
transferred into the area.
In such an area, which is more positively charged than other areas,
the subsequent deposition of Ag will be suppressed,
since the sputtering Ag is positively ionized by
passing through the argon plasma used for RF sputtering.
Such a feedback mechanism induced by optical near-fields
will lead to the unique and nontrivial granular structure
of the Ag film and such a self-controlled pattern formation
was indeed observed in experiments \cite{Yukutake10}.


Naruse et al. \cite{Naruse11} introduced a stochastic
cellular automaton model on a square lattice \cite{CD98}
such that each cell has one of the two values 0 and 1,
which represents a vacancy and an occupation by an Ag grain,
respectively.
Time evolution of Ag clusters in random deposition of Ag$^+$ grains
on a lattice has been numerically simulated.
It was assumed in their model that
the repulsive force acting on an Ag$^+$ grain
when it is deposited on a site in the lattice 
is simply proportional to the total number of occupied
sites in the eight neighbors
({\it i.e.} the Moore neighborhood in the square lattice).
This model works well as a minimal model
for the inhomogeneous pattern formation of Ag film
on an electrode of the photovoltaic device,
while a role of spatially inhomogeneous charge-distribution
due to the heterogeneous effect of 
optical near-fields was not clarified.

In the present paper, we propose a nonequilibrium
statistical-mechanics model on an $L \times L$ square lattice $\Lambda_L$,
in which two kinds of stochastic variables at each site $\r$
evolve in discrete time $t \in \{0,1,2, \dots\}$; 
the number of deposited grain on the site $n_t(\r)$
and the amount of charge per site $q_t(\r)$.
These two variables are dynamically coupled as explained below.
It is known that the effective potential of optical near-field is
well-described by a Yukawa function of a distance $r$,
$V_{\rm eff}(r)=e^{-r/\alpha}/r$.
Here the interaction range $\alpha$ is proportional to the size of the matter
which generates the optical near-field under light illumination \cite{Ohtsu14}.
In the present situation, the range $\alpha$ will be approximately equal
to the size $s$ of Ag-clusters.
Hence growth of sufficiently large clusters is required 
in order to realize the situation such that the optical near-fields generated on 
the Ag surface reach the p-n junction and electron-hole pairs
are created (see Fig.\ref{fig:device2}). 
As a result, the charge increment of a cluster of Ag-deposited sites
depends on the cluster size $s$, which is a functional of $\{n_t(\r)\}_{\r \in \Lambda_L}$.
We introduce a characteristic size $s_0$
of the cluster at which the charge increment of a cluster caused by
the optical near-fields is maximized.
Moreover, we assume existence of a characteristic size-variation $\sigma^2$
such that if the cluster size $s$ deviates from $s_0$ by much larger than
$\sigma$, that is, if $(s-s_0)^2 \gg \sigma^2$, 
then the effect of the optical near-fields for charging cluster
becomes very small.
These assumptions are due to the basic property of DPPs,
the optical near-fields coupled with phonons, called
the {\it size-dependent resonance}, and
here we use the Gaussian function
\begin{equation}
f(s)=\exp \left[ - \frac{(x-s_0)^2}{2 \sigma^2} \right]
\label{eqn:f2}
\end{equation}
to represent such size dependence 
(see Fig.2.6 in Sec.2.2.2 of \cite{Ohtsu14}). 
Furthermore, in our numerical simulation, 
we perform exact calculation of 
long-ranged repulsive Coulomb interaction
between the Ag$^+$ grain being deposited on a site
and positively charged Ag clusters formed on the surface.

The main result of our numerical simulation of the model
is that, when the lattice size $L$ is large but finite, 
there are following two distinct mesoscopic states.
\begin{description}
\item{State A :} \,
If Ag clusters on the surface are formed with a proper size $s$
such that $(s-s_0)^2 \leq \sigma^2$, then the optical near-fields
are induced effectively to increase the charge of clusters 
and hence the total positive charge
on the surface increases monotonically in time.
It causes strong suppression of further deposition of Ag$^+$ grains on the surface
and then the deposition will be stopped.
As a result, a nontrivial and unique Ag film is formed.
\item{State B :} \,
If sizes of Ag clusters on the surface 
tend to deviate from
$s_0$ by more than $\sigma$,
the optical near-fields are not induced effectively.
In this case the charge on surface will be saturated at a low level, 
and the suppression of deposition of Ag$^+$ grains on the surface
remains weak.
Hence the random deposition process will continue without any
self-control and Ag film formation is rather trivial;
as long as laser light is irradiated
the density of Ag grains on the surface gradually increases.
\end{description}
We will show that
if we change some parameters of model,
for instance, a parameter $b$ which denotes
the incident light power per site,
then there occurs a transition 
from State B to State A, and then the self-controlled
formation of Ag file is realized.
We remark that in the real experiments
the p-n junction was reversely biased with a fixed voltage ($V_b=-1.5$ V).
Therefore, the positive charge on the Ag surface is bounded and it will be saturated.
In order to simplify the model, however, we do not take into account
such an effect in the present study.

The paper is organized as follows.
In Sec.\ref{sec:model} the setting of the present model,
the algorithm of processes, and the quantities
which we will calculate are explained.
The results of numerical simulations are given in 
Sec.\ref{sec:results}.
There properties of the dynamical transition of surface state
are studied by the analogy of equilibrium phase transitions
associated with critical phenomena \cite{MD99,MKL09,CD98}.
Sec. \ref{sec:remarks} is devoted to concluding remarks.

\SSC{Discrete-Time Stochastic Model on a Lattice}
\label{sec:model}

Let $L$ be an integer and consider an $L \times L$
square lattice $\Lambda_L=\{1,2, \dots, L\}^2$.
For two sites $\r=(x, y)$, $\r'=(x', y') \in \Lambda_L$ 
with distance $|\r-\r'| \equiv \sqrt{(x-x')^2+(y-y')^2}=1$, 
we say that they are the nearest-neighbor sites. 
We consider a stochastic process with a discrete time
$t \in \N_0 \equiv \{0,1,2, \dots\}$.
Here a particle will represent a positively charged Ag grain.
At each time $t \in \N_0$, the following two kinds of
stochastic variables will be defined, 
\begin{eqnarray}
n_t(\r) &=& \mbox{the number of deposited particles at site $\r$},
\nonumber\\
q_t(\r) &=& \mbox{the amount of charge at site $\r$},
\label{eqn:n_q}
\end{eqnarray}
$\r \in \Lambda_L$, where
$n_t(\r) \in \N_0$ and $q_t(\r) \in \R_+ \equiv \{x \in \R: x \geq 0\}$
(the set of nonnegative real numbers).
At each time $t \in \N_0$, a collection of pairs of
these stochastic variables over the lattice gives
a {\it configuration} of the process, which is denoted by
\begin{equation}
\xi_t=\{(n_t(\r), q_t(\r)) : \r \in \Lambda_L \}, \quad t \in \N_0.
\label{eqn:xi_t}
\end{equation}
In a given configuration $\xi_t$, if $n_t(\r) \geq 1$
and $n_t(\r') \geq 1$ for a pair of the nearest-neighbor sites
$(\r, \r')$, we say that these two sites are {\it connected}.
Moreover, for a pair of sites $\r$ and $\r'$ with
$|\r-\r'| > 1$, if there is a sequence of sites
$\r_0 \equiv \r, \r_1, \dots, \r_{n-1}, \r_n \equiv \r'$ in $\Lambda_L$ with
$|\r_j-\r_{j-1}|=1, 1 \leq j \leq n$, $n \in \{2,3, \dots\}$ such that
any pair of successive sites $(\r_{j-1}, \r_j), 1 \leq j \leq n$ are
connected, then we also say that $\r$ and $\r'$ are {\it connected}.
For each site $\r$ with $n_t(\r) \geq 1$,
a two-dimensional {\it cluster} including the site $\r$ is defined by
\begin{equation}
C_t(\r)=\{\r' \in \Lambda_L :
\mbox{$\r$ and $\r'$ are connected} \}.
\label{eqn:C_t}
\end{equation}
By definition, if $\r$ and $\r'$ are connected,
then $C_t(\r)=C_t(\r')$. 
The total number of sites $\r'$ included in $C_t(\r)$ is
denoted by $|C_t(\r)|$, which represents an area on $\Lambda_L$
of the cluster of deposited particles.

We introduce the parameters in $\R_+$ as
\begin{eqnarray}
b &=& \mbox{incident light power per site},
\nonumber\\
a &=& \mbox{effective coupling constant of}
\nonumber\\
&& \mbox{a repulsive Coulomb potential between charges},
\nonumber\\
E_{\rm th} &=& \mbox{threshold energy for deposition on the surface}.
\label{eqn:parameters}
\end{eqnarray}
If the kinetic energy of Ag$^{+}$ grain injected to the surface
by RF sputtering is relatively small,
the deposition on the surface will be inhibited by the 
Coulomb repulsive force between the Ag$^{+}$ grain and 
the positively charged surface.
In the present stochastic model, we do not calculate
such electrodynamical processes and simply introduce
two variables $a$ and $E_{\rm th}$ as parameters.
We set a function $f: \N_0 \mapsto \R$,
which specifies the increment ratio of charge 
per site and per time-step as a function of cluster size
to which the site belongs.
We assume that $f$ is given by (\ref{eqn:f2}) 
with positive constants $s_0$ and $\sigma$.
Now we explain the elementary process,
$\xi_t \mapsto \xi_{t+1}, t \in \N_0$.

Assume that a configuration $\xi_t$ is given at time $t \in \{0,1,2, \dots\}$.
\begin{description}
\item{(i)} \quad
Choose a site randomly in $\Lambda_L$.
The chosen site is denoted by $\x$.
\item{(ii)} \quad
Calculate the repulsive Coulomb potential at $\x$
caused by all charges on $\Lambda_L$ by
\begin{equation}
V(\x) = \sum_{\r \in \Lambda_L} 
a \frac{q_t(\r)}{|\x-\r|_+},
\label{eqn:Kx}
\end{equation}
where $|\x|_+=|\x|$, if $|\x| >0$ and $|\x|_+=1$, if $|\x|=0$.
(Here we consider a `coarse grain model', and thus
any singularity of the Coulomb-potential function
should be eliminated.) 
\item{(ii-1)} \quad
If $V(\x) \leq E_{\rm th}$, then
a charged particle is deposited at the site $\x$; 
\begin{equation}
n_{t+1}(\r)= \left\{ \begin{array}{ll}
n_t(\r)+1, \quad & \r=\x,
\cr
n_t(\r), \quad & \r \not= \x.
\end{array} \right.
\label{eqn:step1}
\end{equation}
\item{(ii-2)} \quad
If $V(\x) > E_{\rm th}$, then let $\y_0=\x$
and calculate $V(\y)$ at every nearest-neighbor site of $\y_0$, 
$\y \in \cN(\y_0) \equiv \{\y \in \Lambda_L: |\y-\y_0|=1\}$.
If the site which attains $\min_{\y \in \cN(\y_0)} V(\y)$
is uniquely determined, let that site be $\y_1$.
When the minimum is attained by plural sites, we choose
one of them randomly and let it be $\y_1$.
If $V(\y_1) > V(\y_0)$, then 
we regard that the particle is repulsed from the
charged surface and cannot be deposited.
Hence the configuration is not changed at all, 
\begin{equation}
n_{t+1}(\r)=n_t(\r), \quad ^{\forall}\r \in \Lambda_L.
\label{eqn:step2}
\end{equation}
If $V(\y_1) \leq E_{\rm th}$, we set $\x^*=\y_1$.
If $E_{\rm th} < V(\y_1) \leq V(\y_0)$, 
calculate $V(\y)$ at every $\y \in \cN(\y_1)$
and follow the same procedure as above to determine a site $\y_2$.
If $V(\y_2) > V(\y_1)$, 
the deposition is not done and
we have (\ref{eqn:step2}).
If $V(\y_2) \leq E_{\rm th}$, set $\x^*=\y_2$.
If $E_{\rm th} < V(\y_2) \leq V(\y_1)$, we repeat the similar procedure
to those explained above.
When $\x^* \in \Lambda_L$ is determined, 
the charged particle is deposited at the site $\x^*$;
\begin{equation}
n_{t+1}(\r)= \left\{ \begin{array}{ll}
n_t(\r)+1, \quad & \r=\x^*,
\cr
n_t(\r), \quad & \r \not= \x^*.
\end{array} \right.
\label{eqn:step3}
\end{equation}
Each sequence $\x=\y_0 \to \y_1 \to \cdots \to \y_n=\x^*$
with some $n \in \{1,2, \dots\}$
simulates a path of 
{\it drift process} on the surface
performed by a particle
before it is deposited at $\x^*$
\item{(iii)} \quad
Clusters at time $t+1$, 
$\{C_{t+1}(\r) : \r \in \Lambda_L\}$
are redefined for the configuration
$\{n_{t+1}(\r): \r \in \Lambda_L \}$
given by (ii).
For each cluster $C_{t+1}(\r)$,
the accumulated total charge
has been $\sum_{\r' \in C_{t+1}(\r)} q_t(\r')$.
In addition to that, the following amount of charge is
added;
$b |C_{t+1}(\r)| f(|C_{t+1}(\r)|)$.
(See Eq.(\ref{eqn:f2}) and the explanation given above it.) 
Then the charge at site $\r$ at time $t+1$ is given by
\begin{equation}
q_{t+1}(\r) = \frac{1}{|C_{t+1}(\r)|}
\left\{ \sum_{\r' \in C_{t+1}(\r)} q_t(\r')
+ b |C_{t+1}(\r)| f(|C_{t+1}(\r)|) \right\},
\quad \r \in \Lambda_L.
\label{eqn:step_q}
\end{equation}
We note that the charge distribution in cluster can be
spatially inhomogeneous in general depending on the shape of cluster
on the surface. We have assumed the uniform distribution 
in (\ref{eqn:step_q}), however,
since keeping the model simple 
we would like to report the advantage of introducing
the variable $q_t(\r)$, which was not considered
in the previous cellular automaton model \cite{Naruse11},
in order to realize transitions between the two distinct
particle-deposition states.
\end{description}

We start at the empty configuration
$\xi_0=\{(n_0(\r), q_0(\r))=(0,0) : \r \in \Lambda_L\}$
and observe the following quantities at times
$t \in \{0, 1, 2, \dots, T\}$ with a sufficiently large $T$,
\begin{eqnarray}
A_t &=& \sum_{\r \in \Lambda_L} \1(n_t(\r) \geq 1),
\nonumber\\
Q_t &=& \sum_{\r \in \Lambda_L} q_t(\r),
\label{eqn:AQ}
\end{eqnarray}
where $\1(\omega)$ is an indicator function of
a condition $\omega$;
$\1(\omega)=1$ if $\omega$ is satisfied,
and $\1(\omega)=0$ otherwise.
The quantity $A_t$ shows the {\it total area of clusters} at time $t$
and $Q_t$ the {\it total charge} of particles deposited on the surface $\Lambda_L$
at time $t$.
We set
\begin{equation}
R_t= \frac{A_t}{L^2}, 
\label{eqn:R}
\end{equation}
which denotes the {\it occupation ratio} by deposition on the surface
at time $t$.
We also define 
\begin{equation}
Q_t^{(T)}= \frac{Q_t}{Q_T}, \quad t \in \{0, 1,2, \dots, T\}
\label{eqn:QT}
\end{equation}
with a given $T$.

\SSC{Simulation Results}
\label{sec:results}

We performed computer simulation of the stochastic model.
Here we fix the following parameters as
\begin{equation}
a = 1 \times 10^{-3}, \quad s_0=12, \quad
\sigma= 2,
\label{eqn:parameters1}
\end{equation}
and 
\begin{equation}
L=64, \quad T=10^4.
\label{eqn:parameters2}
\end{equation}
The parameters $b$ and $E_{\rm th}$ are changed and dependence
of the process on them is studied.
(We will also discuss the results on systems with different sizes $L$
in Sec.\ref{sec:phase} and Sec.\ref{sec:remarks}.)

\subsection{Two states}
\begin{figure}
\includegraphics[width=1.0\linewidth]{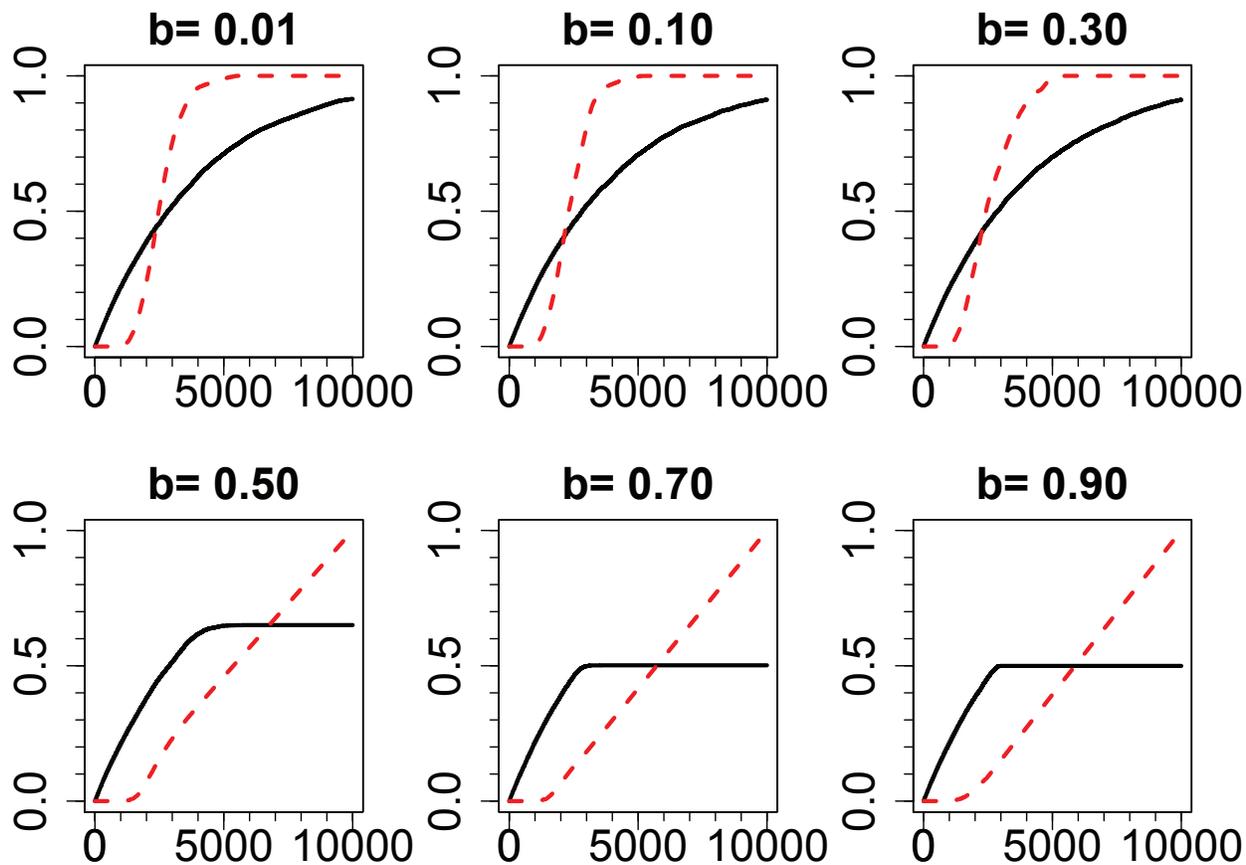}
\caption{$R_t$ (solid lines) and $Q_t^{(T)}$ (broken lines) are shown
as functions of time-steps $t \in \{0,1, \dots, T\}$ 
with $T=10^4$
for $b=0.01, 0.10, 0.30, 0.50, 0.70$, and $0.90$,
where we set $E_{\rm th}=10$.
The abscissas show time steps.
}
\label{fig:RQ_b}
\end{figure}

Figure \ref{fig:RQ_b} shows time dependence
of the occupation ratios by deposition on the surface, $R_t$ (solid lines), 
and the total charges on the surface normalized by
the values at $t=T$, $Q_t^{(T)}$ (broken lines), 
for $b=0.01, 0.10, 0.30, 0.50, 0.70$, and $0.90$,
when we set $E_{\rm th}=10$.
When $b$ is small ($b=0.01, 0.10$ and 0.30), 
$R_t$ increases monotonically and 
$Q_t^{(T)}$ shows saturation after some time-period.
When $b$ is large ($b=0.50-0.90$), on the other hand,
$R_t$ shows saturation in time, while
$Q_t^{(T)}$ increases monotonically.
The results implies when $b$ is small
the system is in State B, while $b$ is large
in State A.
The critical value of $b$ will be evaluated 
in the next subsection as
$b_{\rm c}=0.44$ in this case with $E_{\rm th}=10$.

\subsection{Order parameter and critical exponent}
\label{sec:rho}

When the system is in State B, $R_t$ increases 
monotonically and $\lim_{t \to \infty} R_t \simeq 1$,
that is, the surface will be almost covered by particles.
On the other hand, in State A, the deposition is
suppressed by $Q_t$ which increases monotonically in time
and the surface formation is self-controlled.
As a result we will have a nontrivial steady state, in which
$\lim_{t \to \infty} R_t = R_{\infty} < 1$.

The vacant-site density on the surface in the steady state
is defined as
\begin{equation}
\rho_0 \equiv 1-R_{\infty}. 
\label{eqn:order1}
\end{equation}
It will play a role as an {\it order parameter}
for the transition from State B to State A.

\begin{figure}
\includegraphics[width=0.5\linewidth]{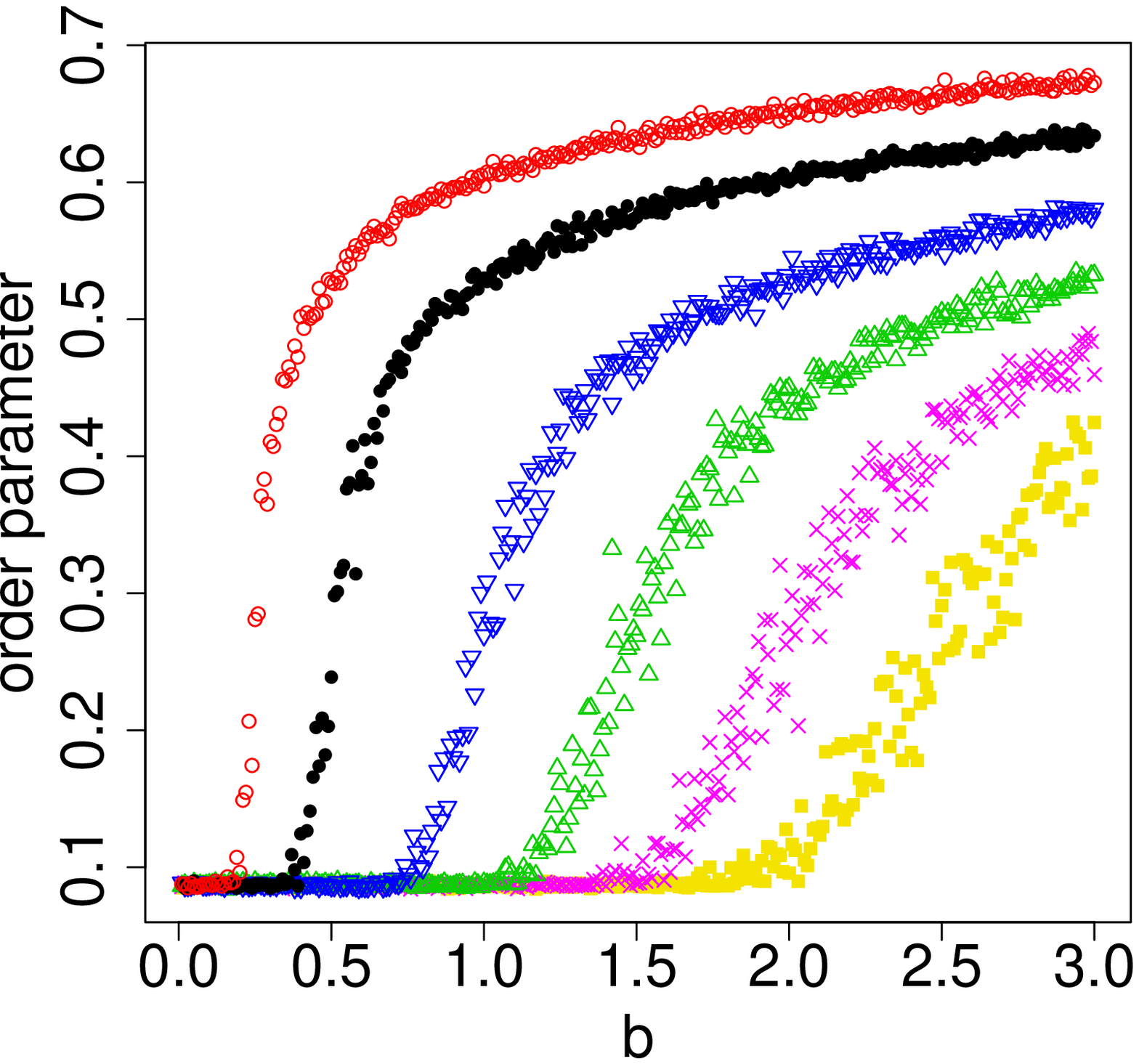}
\caption{Order parameters as functions of $b$
for $E_{\rm th}=5$ (denoted by $\circ)$, $10 \, (\bullet)$, 
$20 (\bigtriangledown)$, 
$30 (\bigtriangleup)$, 
$40 (\times)$, and $50 (\Box)$.
}%
\label{fig:order_parameter}
\end{figure}

Here $\rho_0$ is approximated
by the value 
$\rho_0^{(T)} \equiv 1- R_T$ at $T=10^4$.
For each choice of parameters, we performed
ten independent simulations and studied the averaged values
of $\rho_0^{(T)}$. 
Figure \ref{fig:order_parameter} shows dependence of 
$\rho_0^{(T)}$ on $b$ for
$E_{\rm th}=5, 10, 20, 30, 40$,  and 50.
There $\rho_0^{(T)}$'s seem to behave as continuous functions
of $b$. It implies that the transition
is continuous and the critical value $b_{\rm c}=b_{\rm c}(E_{\rm th})$
will be defined by
\begin{eqnarray}
b_{\rm c}(E_{\rm th})
&=& \max \{b > 0: \rho_0(E_{\rm th}, b) =0 \},
\nonumber\\
&=& \min \{b > 0: \rho_0(E_{\rm th}, b) >0 \}.
\label{eqn:bc1}
\end{eqnarray}

\begin{figure}
\includegraphics[width=0.5\linewidth]{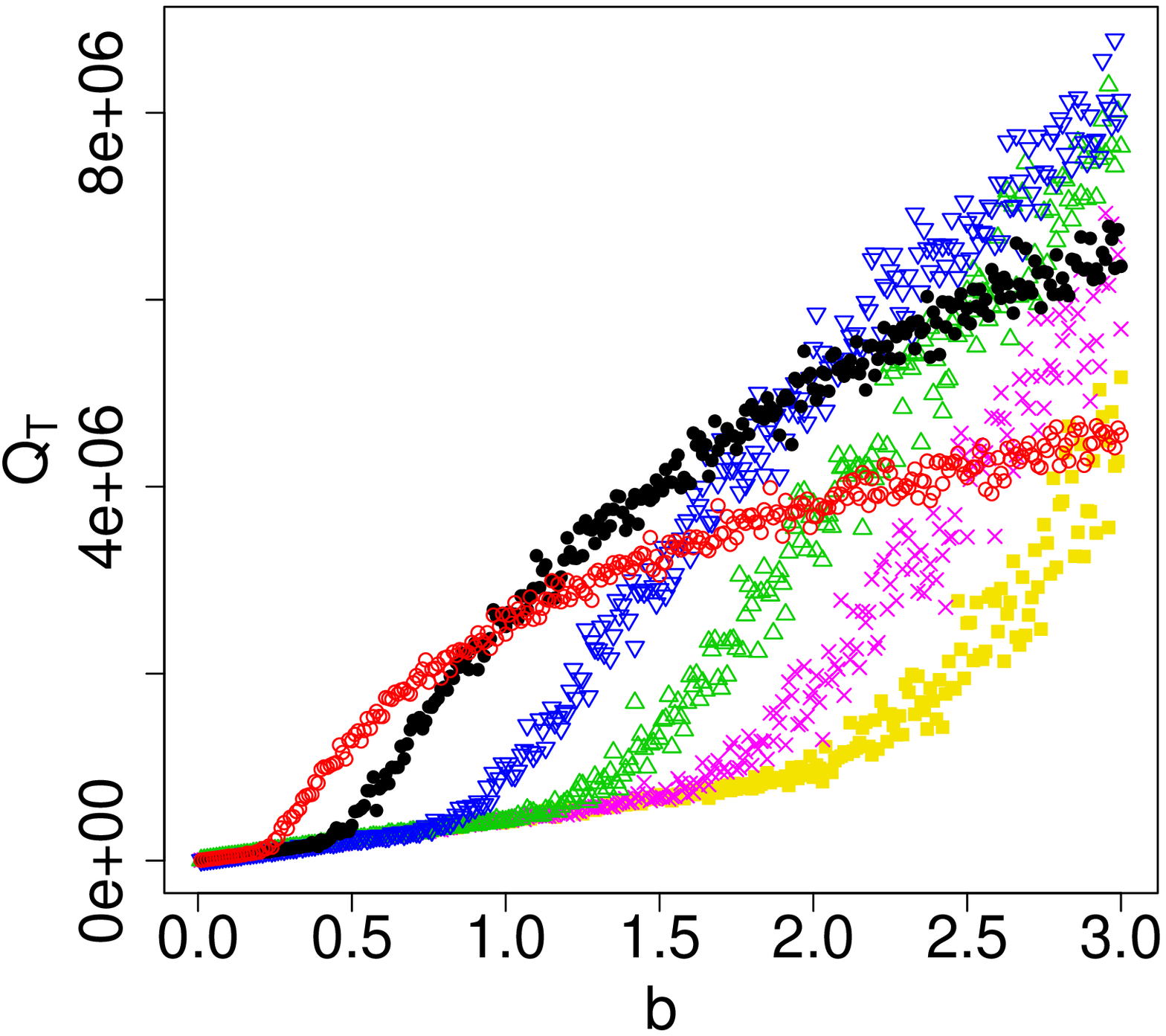}
\caption{$Q_T$ with $T=10^4$ versus $b$ 
for  $E_{\rm th}=5$ (denoted by $\circ)$, $10 \, (\bullet)$, 
$20 (\bigtriangledown)$, 
$30 (\bigtriangleup)$, 
$40 (\times)$, and $50 (\Box)$.
}
\label{fig:Qt_2}
\end{figure}
\begin{figure}
\includegraphics[width=0.5\linewidth]{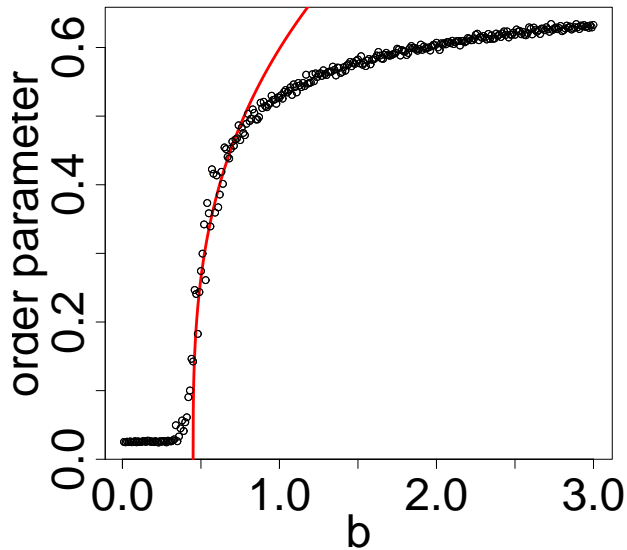}
\caption{Simulation result of the order parameter
$\rho_0$ for $E_{\rm th}=10$.
The curve shows (\ref{eqn:beta1}) 
with $b_{\rm c}=0.44$ and $\beta=0.34$.
In the vicinity of critical value $b_{\rm c}$ in State A,
the power-law behavior (\ref{eqn:beta1})
is observed.
}
\label{fig:fitting}
\end{figure}

Figure \ref{fig:Qt_2} shows the $b$-dependence of 
$Q_T=Q_T(E_{\rm th}, b)$ for $E_{\rm th}=5, 10, 20, 30, 40$, and 50.
The result implies that,
for $b < b_{\rm c}(E_{\rm th})$,
$\lim_{t \to \infty} Q_t < \infty$,
while for $b > b_{\rm c}(E_{\rm th})$,
$\lim_{t \to \infty} Q_t =\infty$
(see the remark given at the end of Sec.\ref{sec:introduction}).
Thus the critical value $b_{\rm c}(E_{\rm th})$
is also characterized as
\begin{eqnarray}
b_{\rm c}(E_{\rm th})
&=& \max \left\{b > 0: \lim_{t \to \infty} Q_t < \infty \right\},
\nonumber\\
&=& \min \left\{b > 0: \lim_{t \to \infty} Q_t = \infty \right\}.
\label{eqn:bc2}
\end{eqnarray}

As an analogy of the second-order phase transitions
associated with critical phenomena
in the equilibrium statistical mechanics, 
singular behavior of the order parameter in the vicinity
of transition point $b_{\rm c}$ in State A
is expected to be described by the following power-law,
\begin{equation}
\rho_0 \sim (b-b_{\rm c})^{\beta}, \quad
0 < b-b_{\rm c} \ll 1,
\label{eqn:beta1}
\end{equation}
with an exponent $\beta$ \cite{MD99,MKL09,CD98}.

For each value of $E_{\rm th}$, we performed the
numerical fitting of the data to the power-law (\ref{eqn:beta1})
and evaluated values of $b_{\rm c}$ and $\beta$.
For example, the curve in Fig. \ref{fig:fitting} shows
the obtained curve (\ref{eqn:beta1}) 
by fitting the data for $E_{\rm th}=10$,
which gives $b_{\rm c}=0.44$ and $\beta=0.34$.
As we will report in the next subsection, 
the critical value $b_{\rm c}$ depends on $E_{\rm th}$,
but we found that the dependence of evaluated $\beta$
on $E_{\rm th}$ is very small;
$\beta=0.31$-0.39 for $E_{\rm th}=5$-50.

\subsection{Critical line between State A and State B}
\label{sec:phase}
\begin{figure}
\includegraphics[width=0.5\linewidth]{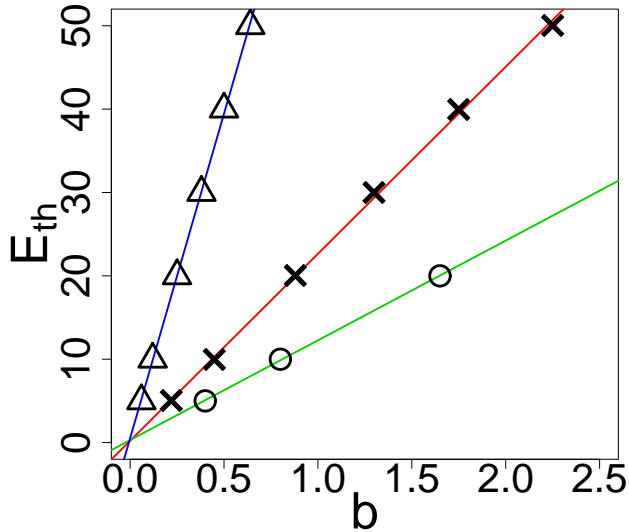}
\caption{
The values of $b_{\rm c}$ of several $E_{\rm th}$'s are
shown by $\times$'s for the system with size $L=64$.
The critical line is well described by a straight line,
$b_{\rm c}=c(L) E_{\rm th}$ with
$c(L=64) \simeq 0.045$,
which divides State A (the lower right region) from
State B (the upper left region).
The critical values evaluated for the systems
with sizes $L=50$ and $L=100$ are
also plotted by $\bigcirc$'s and $\bigtriangleup$'s, respectively.
We have evaluated the coefficients of critical lines as
$c(50) \simeq 0.084$ and $c(100) \simeq 0.013$.
}
\label{fig:phase}
\end{figure}

We have evaluated the critical values $b_{\rm c}=b_{\rm c}(E_{\rm th})$ 
as 
$b_{\rm c}(5)=0.22, b_{\rm c}(10)=0.44,
b_{\rm c}(20)=0.90, b_{\rm c}(30)=1.30, b_{\rm c}(40)=1.75$,
and $b_{\rm c}(50)=2.25$, respectively.
They are plotted by cross marks in Fig. \ref{fig:phase}.  
We can see that 
the {\it critical line} is well described by a straight line,
\begin{equation}
b_{\rm c}= c E_{\rm th}
\quad \mbox{with $c \simeq 0.045$}.
\label{eqn:bvsE}
\end{equation}

As discussed in Sec.\ref{sec:remarks},
the critical value $b_{\rm c}(E_{\rm th})$ also depends on the lattice size $L$.
Because of the long-ranged Coulomb potential (\ref{eqn:Kx}),
$b_{\rm c}(E_{\rm th})$ decreases monotonically as $L$ increases.
In other words, the coefficient in (\ref{eqn:bvsE})
is $L$-dependent; $c=c(L)$, and
$\lim_{L \to \infty} c(L)=0$.
In this sense, the transitions between State A and State B are
not usual phase transitions, which should be defined in 
infinite-size (thermodynamic) limits.
Therefore, Fig. \ref{fig:phase} cannot be regarded as 
a phase diagram.
In the present study, we are interested in transitions
in nonequilibrium states found
on a surface of device with a mesoscopic scale.
The values themselves of $b_{\rm c}(E_{\rm th})$ as well as $c(L)$
reported here are not so important, since they 
change depending on system size.
Our preliminary study by numerical simulations for different lattice sizes imply, however, that 
the linear dependence of $b_{\rm c}$ on $E_{\rm th}$ is
universal as far as the system size is large but finite
as shown in Fig. \ref{fig:phase}.
Note that validity of the power-law (\ref{eqn:beta1})
has been numerically confirmed also for
different sizes of $L$ with $\beta \simeq 0.3$.

\subsection{Cluster structures in State A}

\begin{figure}
\includegraphics[width=0.5\linewidth]{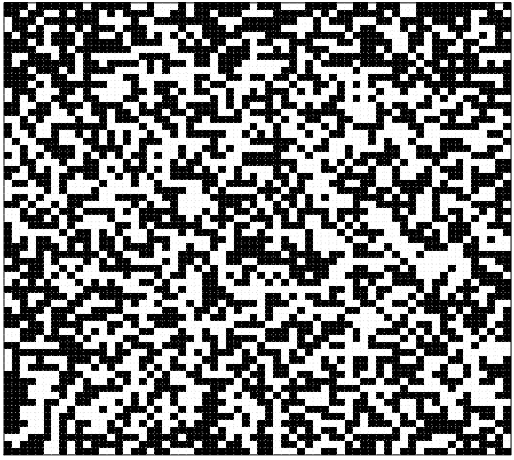}
\caption{Cluster structure formed in the steady state,
when $(E_{\rm th}, b)=(10, 0.65)$.
Occupied sites are dotted.
}
\label{fig:surface}
\end{figure}
\begin{figure}
\includegraphics[width=0.5\linewidth]{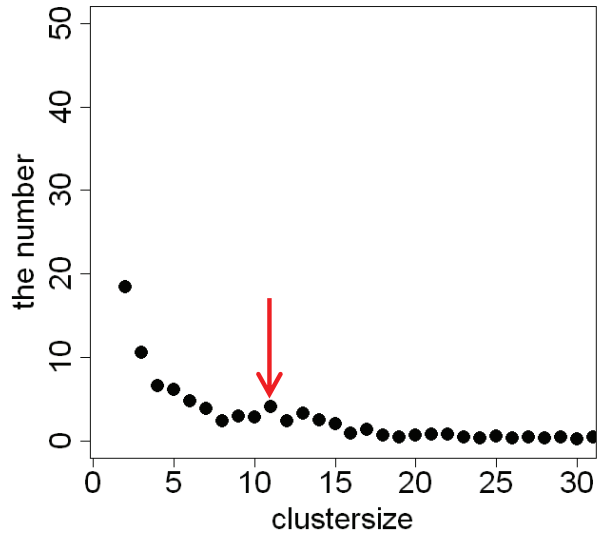}
\caption{Cluster-size distribution in the steady state,
when $(E_{\rm th}, b)=(10, 0.65)$.
As indicated by an arrow, a peak is found around $s \simeq 11$.
}
\label{fig:distribution}
\end{figure}

For $(E_{\rm th}, b)=(10, 0.65)$, 
a configuration of clusters on the surface 
obtained in the steady state (at $T=10^4$)
is plotted in Figure \ref{fig:surface},
where occupied sites by particles are dotted.
For $E_{\rm th}=10$, we have estimated $b_{\rm c}=0.44$.
Then this figure shows a self-controlled surface-configuration
realized in State A.
Averaging over ten independent simulations,
distribution of cluster size $|C_T(\r)|$ on the steady-state surface in State A
is calculated.
Figure \ref{fig:distribution} shows it for $(E_{\rm th}, b)=(10, 0.65)$.
A peak is found around $s \simeq 11$, which is consistent
with our choice of parameter $s_0=12$ in the simulation.
The present results are consistent with the experimental
observation reported by \cite{Yukutake10} as well as
the simulation results by \cite{Naruse11}.

\SSC{Concluding Remarks}
\label{sec:remarks}

In the present paper we generalized the cellular automaton model \cite{CD98}
by Naruse et al. \cite{Naruse11} and proposed a new stochastic model
to describe the nonequilibrium dynamics of Ag-film pattern formation
on an electrode of the photovoltaic device of Yukutake et al. \cite{Yukutake10}.
In the present new model, not only the number of deposited particle
$n_t(\r)$ but also the amount of charge $q_t(\r)$ are considered
as stochastic variables at each site $\r$ in the $L \times L$
square lattice at time $t \in \N_0$.
The previous model used the `pseudo footprint' method
to take into account the repulsive interaction
between an Ag$^+$ ion approaching the surface and 
positively charged Ag clusters on the surface.
In the present algorithm of model, the repulsive Coulomb
potential at each site caused by all charges on the lattice except the site
is exactly calculated at each time-step, 
and drift and deposition processes are simulated.
The essential improvement of the model is such that
we include the effect of optical near-fields generated by
light-matter interaction by introducing a parameter $b$
and a function $f(s)$. There $b$ denotes the
incident light power per site on the lattice
and $f(s)$ expresses the size-dependent resonance effect
of optical near-fields on matter-size $s$.
We have regarded $b$ as an external control parameter of the model.
We have shown that, 
as $b$ increases, there occurs a transition
of the surface state of pattern formation with
a critical value $b_{\rm c}$, such that
if $b \leq b_{\rm c}$ the optical near-fields are
not induced effectively (State B), while if $b > b_{\rm c}$
they are induced effectively and Ag deposition process
is self-controlled (State A). A nontrivial and
unique Ag film is formed in State A,
while in State B random Ag-deposition will be continued
as long as the simulation is continued.
We have discussed the transitions of surface processes
from State B to State A by the analogy of
equilibrium second-order phase transitions
with critical phenomena \cite{MD99,MKL09,CD98}.
We also studied the dependence of $b_{\rm c}$
on the threshold energy $E_{\rm th}$ for elementary 
deposition of single Ag$^{+}$ on the surface.

The transition of surface state discussed in this paper is
not a phase transition in the usual sense, since our model
is defined on a lattice with a finite size $L < \infty$; $L=64$ for
the simulation data reported above.
In equilibrium and hence in stable systems
the $L \to \infty$ limit will provide the thermodynamic limit
and it describes macroscopic behavior of materials.
In the present model, however, the direct $L \to \infty$ limit
may be meaningless, since the repulsive Coulomb repulsive 
potential (\ref{eqn:Kx}) will diverge as $L \to \infty$
in the late stage of process in which sufficient amount of charge 
is accumulated on the surface.
(It implies that $b_{\rm c}$ should be a decreasing
function of $L$, since $V(\x)$ given by (\ref{eqn:Kx})
will take larger value as $L$ increases,
and $b_{\rm c} \to 0$ as $L \to \infty$
for all $E_{\rm th} > 0$.)
In the present study, we are interested in 
nonequilibrium deposition dynamics of charged particles
on mesoscopic materials.
We should assume that the system size $L$ of model
is sufficiently large but finite.

From the view point of study of 
nonequilibrium statistical mechanics \cite{MD99,MKL09,CD98},
it is an interesting problem to find a relevant
scaling limit which can realize the present
transitions of surface state as nonequilibrium phase transitions.
A preliminary study by changing the lattice size $L$
in numerical simulations implies that
the parameter $b$ should be scaled as
$b=b_0 L^{-(2+\nu)}$ with a constant $b_0$ and
a scaling exponent $\nu$ in the infinite-size limit $L \to \infty$.
The dependence of $c(L)$ on $L$ reported in Fig. \ref{fig:phase} is 
consistent with it and gives a preliminary estimate
$\nu \simeq 0.7$.
Such a mathematical physics aspect of the present model
is also interesting and 
will be reported elsewhere.

\vskip 0.5cm
\noindent{\bf Acknowledgements} \quad
The present authors would like to thank
S. Tojo for useful discussions on the present work.
This study is supported by the Grant-in-Aid for Challenging Exploratory
Research (No.15K13374) of Japan Society for the Promotion of Science. 
MK is supported in part by
the Grant-in-Aid for Scientific Research (C)
(No.26400405) of Japan Society for
the Promotion of Science.




\begin{thebibliography}{}\label{sec:TeXbooks}

\bibitem{Ohtsu08}
M. Ohtsu, T. Kawazoe, T. Yatsui, and M. Naruse, 
IEEE J. Sel. Top. Quantum Electron. {\bf 14}, 1404 (2008). 

\bibitem{Tojo05}
S. Tojo and M. Hasuo, 
Phys. Rev. A {\bf 71}, 012508 (2005).

\bibitem{Ohtsu14}
M. Ohtsu,
{\it Dressed Photons: Concepts of Light-Matter Fusion Technology},
Nano-Optics and Nanophotonics
(Springer, Berlin, 2014).

\bibitem{Franzl04}
T. Franzl, T. A. Klar, S. Schietinger, A. L. Rogach, and J. Feldmann, 
Nano Lett. {\bf 4}, 1599 (2004).

\bibitem{Yukutake10}
S. Yukutake, T. Kawazoe, T. Yatsui, W. Nomura, K. Kitamura, and M. Ohtsu,
Appl. Phys. B {\bf 99}, 415 (2010).

\bibitem{Pistol08}
C. Pistol, C. Dwyer, and A. R. Lebeck, 
IEEE MICRO {\bf 28}, 7 (2008).

\bibitem{Kawazoe11}
T. Kawazoe, M. A. Mueed, and M. Ohtsu, 
Appl. Phys. B {\bf 104}, 747 (2011).

\bibitem{Naruse13}
M. Naruse, N. Tate, M. Aono, and M. Ohtsu, 
Rep. Prog. Phys. {\bf 76}, 056401 (2013).

\bibitem{Naruse13b}
M. Naruse, ed.,
{\it Nanophotonics Information Physics:
Nanointelligence and Nanophotonics Computing}, 
Nano-Optics and Nanophotonics (Springer, Berlin, 2014).

\bibitem{Yatsui05}
T. Yatsui, W. Nomura, and M. Ohtsu, 
Nano Letters {\bf 5}, 2548 (2005).

\bibitem{MD99}
J. Marro and R. Dickman,
{\it Nonequilibrium Phase Transitions in Lattice Models},
(Cambridge University Press, Cambridge, 1999).

\bibitem{MKL09}
R. Mahnke, J. Kaupu{\v z}s, and I. Lubashevsky,
{\it Physics of Stochastic Processes:
How Randomness Acts in Time},
(Wiley-VCH, Germany, 2009).

\bibitem{CD98}
B. Chopard and M. Droz,
{\it Cellular Automata Modeling of Physical Systems},
(Cambridge University Press, Cambridge, 1998).

\bibitem{Naruse11}
M. Naruse, T. Kawazoe, T.Yatsui, N. Tate, and M. Ohtsu,
Appl. Phys. B {\bf 105}, 185 (2011).

\bibitem{KSIO00}
K. Kobayashi, S. Sangu, H. Ito, and M. Ohtsu, 
Phys. Rev. A {\bf 63}, 013806 (2000).
\end{thebibliography}
\end{document}